\documentclass[12pt]{article}
\newfont{\bbd}{msbm10 scaled\magstep1}
\def\C{\hbox{\bbd C}}
\begin{document}
\renewcommand{\thefootnote}{\fnsymbol{footnote}}
\begin{center} {\Large \bf Kowalevski top revisited}
%\footnote[2]{This work is based on the talk given in the Newton Institute,
%Cambridge, UK, on September 7, 2001}
\end{center}
\renewcommand{\thefootnote}{\arabic{footnote}}
\vskip1cm
\renewcommand{\thefootnote}{\fnsymbol{footnote}}
\begin{center} {\bf Vadim B.~Kuznetsov\footnote[2]{EPSRC 
Advanced Research Fellow}}
\end{center}
\renewcommand{\thefootnote}{\arabic{footnote}}
\vskip 0.5cm
\begin{center} Dept of Applied Maths, University of Leeds, LEEDS LS2 9JT, 
United Kingdom\\
E-mail: vadim@maths.leeds.ac.uk
\end{center}
\vskip2cm
\begin{center}
{\bf Abstract} \end{center}
\vskip0.2cm

\noindent
We review the separation of variables for the Kowalevski top and for its
generalization to the algebra o(4). We notice that the corresponding separation
equations allow an interpretation of the Kowalevski top as a $B_2^{(1)}$ 
integrable 
lattice. Consequently, we apply the quadratic $r$-matrix formalism to construct 
a new $2\times2$ Lax matrix for the top, which is responsible for its
separation of variables.
\vskip 5cm

\noindent
To appear in: Kowalevski property (2002). CRM Proc. \& Lecture Notes, 
(V.~B.~Kuznetsov, Editor), surveys from Kowalevski Workshop
on Mathematical Methods of Regular Dynamics (Leeds, April 2000),
American Mathematical Society
\vskip 5cm

\pagebreak
%
%%%%%%%%%%%%%%%%%%%%%%%%%%%%%%%%%%%%%%%%%
%
\section*{Introduction}
In 1889 Sophie Kowalevski \cite{SK} found and integrated new integrable case of 
rotation of a heavy rigid body around a fixed point which since then 
carried her name, {\it the Kowalevski top} (KT). In modern terms, this
is an integrable system on the e(3) algebra with a quadratic and a quartic
(in angular momenta) integrals of motion.

Her original transformation
to new variables led to the solution in terms of quadratures and,
eventually, to
the separation of variables for this complicated top \cite{VN,Ku85,KK87}. 
No separation which is 
alternative to her original separation of variables is known for this system 
at the moment, even though there is a large body of literature dedicated to the 
problem (see, for instance, \cite{HH,AM88,F88,HM89,BRS,KT89a} to name
just a few).

The aim of the present work is to design and construct a new Lax
matrix for the Kowalevski top which implies Kowalevski's separation 
of variables and which has a proper algebraic ($r$-matrix) structure.
The principal existence of such ($2\times2$) Lax matrix was predicted
in author's MSc dissertation in 1985 \cite{Ku85} (see also \cite{KK87}), 
after he rewrote the formula defining the separation variables for the top 
given in \cite{VN}, and this new formula took a form 
of the spectral curve of the future Lax matrix. In what follows I shall explain
how this Lax matrix can be reconstructed in somewhat regular way, without
much of a guesswork. 

In 1981 the Kowalevski top was generalized to the case of o(4)
algebra \cite{K81}. In 1990 for this o(4) KT the separation of variables
was found in \cite{Ku89,KK90} which generalized the one for the e(3) KT. 
In the present paper we will always consider this one-parameter,
o(4) extension of the Kowalevski top. We refer the reader to 
the work \cite{K00} 
for the survey of the Kowalevski top, its generalizations and other
related tops.

The structure of the paper is following. In Section 1 we recall the 
integration in quadratures of the o(4) Kowalevski top. In Sections 2
and 3 we describe its separation of variables and, in particular, write down 
the separation equations in the form which is best for deducing the ansatz 
for the future Lax matrix. As a result, we conclude that, as it follows
from the separation equations, the Kowalevski top
(together with its o(4) version) is an integrable system with boundary
conditions of $B_2^{(1)}$-type which means that the corresponding
Lax matrix should factorize and factors obey the reflection
equation (quadratic $r$-matrix algebra). In Section 4 we 
introduce the corresponding quadratic algebra ${\cal B}$. Section 5 contains
separation representation for this algebra, while Section 6 defines
and solves the $B_2^{(1)}$-type integrable system. 
In Section 7 the $2\times 2$ Lax matrix $T(u)$ for the o(4) Kowalevski top is given
which follows by comparing the separation data for two integrable
systems from the previous Sections (KT and the integrable system
on the quadratic algebra ${\cal B}$). 
The final result is given in terms of initial variables of the top.

%
%%%%%%%%%%%%%%%%%%%%%%%%%%%%%%%%%%%%%%%%%
%
\section{Integration in quadratures of the o(4) Kowalevski top}
In this Section we collect the results from \cite{Ku89,KK90} about integration in quadratures
of the Kowalevski top on the o(4) algebra.

The Poisson brackets for the o(4) generators $J_k$, $x_k$, $k=1,2,3$, 
are defined in the standard way:
\begin{equation}\label{pb}
\{J_i,J_k\}=\varepsilon_{ikl} J_l, \qquad\{J_i,x_k\}=\varepsilon_{ikl} x_l,\qquad
\{x_i,x_k\}=-{\cal P}\varepsilon_{ikl} J_l,
\end{equation}
where $\varepsilon_{ikl}$ is the completely anti-symmetric tensor, 
$\varepsilon_{123}=1$,
and ${\cal P}$ is a complex (or real) parameter. In fact, one can think of the 
algebra (\ref{pb}) as a complex o(4,\C) algebra with the triple of real algebras
appearing
when the parameter ${\cal P}$ is specialized as follows:
\begin{equation}\label{p}
{\cal P}=\left\{   \matrix{-1,\qquad \mbox{\rm o(4)}\cr \;\;0,\qquad\mbox{\rm e(3)}\cr
                     \;\;\;\;\;1,\qquad \mbox{\rm o(3,1)}}\right..
\end{equation}
The Casimirs of the bracket (\ref{pb}) have the form
\begin{equation}\label{c}
\ell=x_1J_1+x_2J_2+x_3J_3,\qquad a^2=x_1^2+x_2^2+x_3^2-{\cal P}(J_1^2+J_2^2+J_3^2).
\end{equation}

The o(4) Kowalevski top has the Hamiltonian $H$,
\begin{equation}\label{h}
H=J_1^2+J_2^2+2J_3^2-2bx_1,
\end{equation}
and the second integral $K$,
\begin{eqnarray}
K=(J_+^2+2bx_+-{\cal P}b^2)(J_-^2+2bx_--{\cal P}b^2),\label{k}\\
J_\pm=J_1\pm iJ_2,\qquad x_\pm=x_1\pm ix_2, \qquad (i^2=-1),
\nonumber
\end{eqnarray}
which are Poisson commuting:
\begin{equation}\label{0}
\{H,K\}=0.
\end{equation}
When ${\cal P}=0$, this Liouville integrable system
becomes the usual e(3) Kowalevski top. The o(4)
version was proved to be integrable in \cite{K81} 
and it was integrated in \cite{KK90}, where also the separation
of variables was performed for this system, with separation equations being 
put in the form which is best suited for the purpose of this paper.

The integration of the o(4) KT in \cite{KK90} was inspired by the paper 
\cite{HH} about integration of the e(3) case where it was done
by mapping the KT into the Neumann system, thereby giving another way
of looking at the original Kowalevski's transformation.
For the o(4) case it goes as follows. Introduce new variables:
\begin{equation}\label{pl}
p_1=\frac{1+J_+J_-}{2(J_+-J_-)},\qquad 
p_2=\frac{J_++J_-}{2i(J_+-J_-)},\qquad 
p_3=\frac{1-J_+J_-}{2i(J_+-J_-)},
\end{equation}
\begin{equation}\label{pl2}
l_1=\frac{J_3(1-J_+J_-)-bx_3(J_++J_-)}{2(J_+-J_-)},\;\;
l_2=\frac{ibx_3}{J_+-J_-},\;\;
l_3=\frac{J_3(1+J_+J_-)+bx_3(J_++J_-)}{2(J_+-J_-)}.
\end{equation}
Under the flow governed by the Kowalevski Hamiltonian,
\begin{equation}\label{dot}
\dot{(.)}\equiv\frac {d\;(.)}{dt}=\frac12\{H,(.)\},
\end{equation}
these new variables, $p_k$, $l_k$, $k=1,2,3$, combined into
two vectors $\vec p$ and $\vec l$, evolve as the ``e(3) Neumann
system'':
\begin{equation}\label{ns}
\dot{\vec p}=2i\;\vec l\times \vec p,\qquad \dot{\vec l}=2i\; Q\vec p\times \vec p,
\end{equation}
with the symmetric matrix $Q$ depending on Casimirs and both integrals:
\begin{equation}
Q=\pmatrix{-\frac14+b^2\tilde a&ib\ell&i(\frac14+b^2\tilde a)\cr
                  ib\ell&-\frac{H}{2}&-b\ell\cr
                   i(\frac14+b^2\tilde a)&-b\ell&\frac14-b^2\tilde a},
\end{equation}
\begin{equation}\label{aa}
\tilde a=a^2-K/(4b^2)+{\cal P}H/2+b^2{\cal P}^2/4.
\end{equation}

The Neumann system (\ref{ns})--(\ref{aa}) has four integrals of motion:
\begin{equation}
I_1={\vec l}{\;}^2+(Q\vec p,\vec p),\quad
I_2=(Q\vec l,\vec l\;)-(Q^\wedge\vec p,\vec p),\quad
C_1=(\vec l,\vec p)=0, \quad C_2=\vec p{\;}^2=-\frac14,
\end{equation}
where $Q^\wedge$ denotes the adjoint matrix. It is usually
formulated together with the canonical e(3) Poisson bracket
for the generators $p_k$ and $l_k$. Unfortunately, the transformation
(\ref{pl})--(\ref{pl2}) is non-canonical, so it does not bring the o(4) 
bracket of the 
initial variables into this e(3) bracket of the Neumann system.
This means we can use the equations (\ref{ns}) only keeping in mind that we
have a new Poisson structure on the variables $p_k$ and $l_k$ 
imposed by the Kowalevski o(4) Poisson
structure.

As was already mentioned, the mapping of the e(3) Kowalevski dynamics (${\cal P}=0$) 
into Neumann system's dynamics was first found in \cite{HH}. 
Algebraic geometric arguments proving the existence of such
(non-canonical) maps in other `genus 2' situations was
also studied in detail around that time (cf. \cite{AM88} and references therein) 
producing many other maps between
various systems, in particular, tops.

Now, let us proceed further to the integration in quadratures.
Since the problem has been reduced to the dynamics of the Neumann 
system, its integration is well-known (see, for instance, \cite{M}). 
One introduces two new variables $\lambda_1$ and $\lambda_2$
as zeros of the polynomial:
\begin{equation}\label{ll}
\lambda^2+\lambda\left(\frac{H}{2}-4(Q\vec p,\vec p)\right)
-4(Q^\wedge \vec p,\vec p)=
(\lambda-\lambda_1)(\lambda-\lambda_2).
\end{equation}
Then one checks that they satisfy the following equations:
\begin{equation}\label{ke}
\dot{\lambda_1}(\lambda_1-\lambda_2)=2\sqrt{-R_5(\lambda_1)},\qquad
\dot{\lambda_2}(\lambda_2-\lambda_1)=2\sqrt{-R_5(\lambda_2)},
\end{equation}
\begin{equation}\label{ke2}
R_5(\lambda)=\left((\lambda-{\cal P}b^2/2)^2-K/4\right)
\left[\left(\lambda+\frac{H}{2}\right)
                           (\lambda^2+b^2\tilde a)-b^2\ell^2\right].
\end{equation}
After rewriting these equations in the form
\begin{equation}\label{aj}
\frac{d\lambda_1}{\sqrt{-R_5(\lambda_1)}}
+\frac{d\lambda_2}{\sqrt{-R_5(\lambda_2)}}=0,\qquad
\frac{\lambda_1d\lambda_1}{\sqrt{-R_5(\lambda_1)}}
+\frac{\lambda_2d\lambda_2}{\sqrt{-R_5(\lambda_2)}}=2dt,
\end{equation}
one uses Abel-Jacobi map to integrate the dynamics in terms of theta-functions.
This is exactly what Kowalevski did in 1889 (for ${\cal P}=0$): 
she found the variables $\lambda_1$
and $\lambda_2$ (\ref{ll}), she derived the equations (\ref{ke})--(\ref{ke2}), 
now known as 
the Kowalevski equations, and she found theta-function formulas for the 
initial variables of the top, $J_k$ and $x_k$, as functions of time $t$. Of course, she
was not aware of the `simplifying map' to the Neumann system discovered 100 years
later, this is why her original calculations looked so complicated,
mysterious and attractive for many generations of mathematicians,
and they still do!

Formulas (\ref{ll})--(\ref{aj}) for arbitrary ${\cal P}$ were derived in \cite{KK90}.

%
%%%%%%%%%%%%%%%%%%%%%%%%%%%%%%%%%%%%%%%%%
%
\section{Separation of variables for the o(4) Kowalevski top}

In this Section we collect the results from \cite{KK90} about separation of variables
(SoV) for the Kowalevski top on the o(4) algebra.

Separation variables have been generally used to construct analytic
expressions for the action variables (in terms of abelian integrals)
or in order to get a separated representation for the action function.
Therefore, the method of separation of variables for a long time
served an important, but technical role in solving Liouville 
integrable systems of classical mechanics.
A new, and much more exciting, application of the method came with
the development of quantum integrable systems. Because of the fact 
that quantization of the 
action variables seems to be a rather formidable task, quantum separation
of variables became an inevitable refuge. In fact, it was successfully
performed for many families of integrable systems (see, for instance,
survey \cite{Skl}).

We should start from a (working) definition of the SoV. Historically, several
very different definitions have been given, each depending on the context. We
will always mean the context of finite-dimensional integrable Hamiltonian dynamics
or, in other words, our definition will only be valid for the Liouville integrable 
systems. 

By {\it separation of variables} we mean a canonical transformation to new
variables $u_j,v_j$, $j=1,\ldots,n$, which satisfy the following 
{\it separation equations}
\begin{eqnarray}\label{se}
\sum_{j=1}^n a_{ij}H_j=b_i,\qquad i=1,\ldots,n,\\
a_{ij}=a_{ij}(u_i,v_i),\qquad b_i=b_i(u_i,v_i),\label{se-1}
\end{eqnarray}
or, in other words, in terms of which the integrals of motion $H_j$ acquire 
the following `separated form':
\begin{equation}\label{sf}
\vec H=A^{-1}\vec B,\qquad (A)_{ij}=a_{ij},\qquad (\vec B)_i=b_i,\qquad (\vec H)_i=H_i.
\end{equation}
The conditions (\ref{se-1}) that the functions $a_{ij}$ and $b_i$ in (\ref{se}) depend
on the new (separation) variables with the index $i$  {\it only}, is crucial. It indeed
means that the $n$ equations in (\ref{se}) are really separated from one another.

The above definition includes, as a special canonical transform,
the (classical) coordinate separation of variables,
when the new coordinates, say $u_j$, are the functions of the
old coordinates ($q_j$) only, and they do not depend on the momenta ($p_j$). 
See the book \cite{Ka} about 
the history of this sub-class of transformations with many examples
of such situation. 

General (separating) canonical transforms, however, result
in new (separation) variables being non-trivial functions of all $2n$
initial canonical variables,
\begin{eqnarray}\label{ss-exx}
&&u_j=u_j(q_1,\ldots,q_n,p_1,\ldots,p_n),\qquad
v_j=v_j(q_1,\ldots,q_n,p_1,\ldots,p_n),\qquad j=1,\ldots,n,\qquad\\
&&\{u_j,v_k\}=0,\qquad j\neq k,\qquad \{v_j,u_j\}=1,\qquad j=1,\ldots,n.
\end{eqnarray}
The examples of such separating canonical transforms are usually
much more sophisticated than those of the coordinate ones. As far as I know,
the first explicit example was given by van Moerbeke in \cite{M76}
concerning separation of variables for the Toda lattice
(see also \cite{FM}). The method
was further developed by Komarov in a series of works on tops, including
quantum separation of variables, see \cite{Kom,Kom2}. Many further examples 
have been
produced since 1982, with the theory benefiting mostly from the developments
of the algebraic geometric and $r$-matrix understanding of the 
method of separation of variables. This led to a rather satisfying picture
of the present state-of-art of non-coordinate separation of variables.
See \cite{Skl} and \cite{Hu,Hu2} for more details.

It is interesting to remark that many explicit non-coordinate separations
 were already
produced by the classics, such as those for the Goryachev-Chaplygin
and Kowalevski top, {\rm but} they never calculated the Poisson brackets,
so that many modern developments involved checking the canonicity
of  the transformations, proposed by mathematicians of the 19th century!
Hence, classics {\it knew} about non-coordinate separation of variables
(sometimes more than we do now). It is a pity that we lack a good
review of this subject which would cover, say, the separation of variables
for the Clebsch, Euler-Manakov, Steklov and many other tops, including
modern results and those obtained by classics.

Two pairs of separation variables for the o(4) Kowalevski top are as follows:
\begin{equation}\label{ss}
s_i=2\lambda_i+H,\qquad p_i=\frac{1}{2\sqrt{-2s_i}}
\ln \frac{\xi_i+\sqrt{\xi_i^2+d_i^2}}{d_i},
\qquad i=1,2,
\end{equation}
\begin{equation}\label{pp}
\xi_i=2\sqrt{y_i^2+d_iy_i},\quad y_i=(s_i-H-{\cal P}b^2)^2-K,\quad
d_i=4b^2\left(\frac{{\cal P}s_i}{2}+a^2-\frac{2\ell^2}{s_i}\right).
\end{equation}
They are canonical variables,
\begin{equation}\label{ca}
\{s_1,s_2\}=\{p_1,p_2\}=\{s_1,p_2\}=\{s_2,p_1\}=0,
\end{equation}
\begin{equation}\label{ca2}
\{p_1,s_1\}=1,\qquad \{p_2,s_2\}=1, 
\end{equation}
and they satisfy the following separation equations:
\begin{equation}\label{se3}
s_i^3-(2H+{\cal P}b^2)s_i^2+\kappa s_i-4b^2\ell^2=
2b^2\left(\frac{{\cal P}s_i^2}{2}+a^2s_i-2\ell^2\right)
\cos(2\sqrt{2s_i}p_i),
\end{equation}
\begin{equation}\label{se4}
\kappa=(H+{\cal P}b^2)^2-K+2a^2b^2.
\end{equation}
Notice that the role of the $H_1$ and $H_2$ from the definition (\ref{se})
is played here by the integrals $H$ and $\kappa$. The functions $a_{ij}$
and $b_i$ can be directly read from the formulas (\ref{se3}). Also notice
that the formulas (\ref{se3}) are equivalent to two formulas for $p_i$'s
from (\ref{ss}).

This result, i.e. the separation of variables, 
was given for ${\cal P}=0$ in \cite{VN}\footnote{without a proof} 
and for the general case
in \cite{KK90}. We will give here a shorter proof than the one in 
\cite{KK90}, but before
doing that let us reformulate the statement in terms of other variables, 
the $s_i$ and $y_i$. 

The variables $s_i$ and $y_i$, $i=1,2$, are `almost canonical',
\begin{equation}\label{ca3}
\{s_1,s_2\}=\{y_1,y_2\}=\{s_1,y_2\}=\{s_2,y_1\}=0,
\end{equation}
\begin{equation}\label{ca4}
\{y_1,s_1\}=-4\sqrt{-2s_1y_1(y_1+d_1)},\qquad 
\{y_2,s_2\}=-4\sqrt{-2s_2y_2(y_2+d_2)}, 
\end{equation}
and they satisfy the following simple separation equations:
\begin{equation}\label{se5}
y_i=(s_i-H-{\cal P}b^2)^2-K,\qquad i=1,2.
\end{equation}
In order to bring the equations (\ref{ca3})--(\ref{se5}) into the equations
(\ref{ca})--(\ref{se4}) one needs to make a transformation from $y_i$'s
to the canonically conjugated variables $p_i$'s which is easily found
by taking the integral,
\begin{equation}\label{ca5}
p_i=-\frac14\int^{y_i}\frac{dy}{\sqrt{-2s_iy(y+d_i)}}=
\frac{1}{2\sqrt{-2s_i}}\ln\left( 1+\frac{2y_i}{d_i}+2\sqrt{\frac{y_i}{d_i}
\left( \frac{y_i}{d_i}+1 \right)} \,\right).
\end{equation}
The separation equations (\ref{se5}) are then transformed into
the equations
\begin{equation}\label{ca6}
s_i\left( y_i+\frac{d_i}{2}  \right)=\frac{s_id_i}{2}
\cos(2\sqrt{2s_i}p_i),
\end{equation}
which are equivalent to the separation equations (\ref{se3}).
Notice here that $R_5(\lambda_i)$ from (\ref{ke})--(\ref{ke2})
have the following expressions in terms of the variables
$s_i$, $y_i$ and $d_i$:
\begin{equation}\label{caca6}
R_5(\lambda_i)=\frac{s_iy_i}{32}(y_i+d_i),\qquad i=1,2.
\end{equation}

Finally, the brackets (\ref{ca3})--(\ref{ca4}) are checked by a
direct computation\footnote{on a computer: checking (\ref{ca3})
takes about 2 hours}.

%
%%%%%%%%%%%%%%%%%%%%%%%%%%%%%%%%%%%%%%%%%
%
\section{Hyperelliptic Prymian and ansatz for the Lax matrix}

In this Section we change the separation variables $s_i$ and $p_i$
to new separation variables $u_i$ and $m_i^\pm$, which respect 
the symmetry of the problem, and while doing 
that we will derive the proper ansatz for the future $2\times 2$ Lax matrix
of the o(4) Kowalevski top.

First of all, notice that in the definition of $p_i$'s in (\ref{ss}) there appears
$\sqrt{s_i}$, which is very strange taking into account that
 $\lambda=-\frac{H}{2}$ (which is equivalent to $s=0$ according to the 
definition of $s$-variables (\ref{ss})) is {\it not} a branching point 
of the Kowalevski curve $\mu^2=R_5(\lambda)$ (cf. (\ref{ke2})).

This `paradox' was explained in \cite{KK90} by stating that the $s$-variables 
are {\it not} the proper variables and one should introduce new, $u$-variables,
which respect the additional symmetry present in the problem\footnote{
for the usual e(3) Kowalevski top, i.e. in the ${\cal P}=0$ case, the original
explanation was presented much earlier in author's MSc dissertation in 1985, 
see \cite{Ku85}, \cite{KK87}}. This problem of $s$-variables passed unnoticed
in \cite{VN} in the e(3) case.

The proper new variables $u_i$ and $\hat y_i$, $i=1,2$, are introduced as follows:
\begin{equation}\label{ss10}
u_i:=\sqrt{\frac{s_i}{2}},\qquad
\hat y_i:=(u_i^2-H/2-{\cal P}b^2/2)^2-K/4,\qquad i=1,2.
\end{equation}
Another pair of variables, which are functions of $u$-variables and
Casimirs, will be used:
\begin{equation}\label{ss11}
\hat d_i:=b^2\left({\cal P}u_i^2+a^2-\frac{\ell^2}{u_i^2}\right),\qquad i=1,2.
\end{equation}
The variables $u_i$ and $\hat y_i$ are `almost canonical':
\begin{equation}\label{ca13}
\{u_1,u_2\}=\{\hat y_1,\hat y_2\}=\{u_1,\hat y_2\}=\{u_2,\hat y_1\}=0,
\end{equation}
\begin{equation}\label{ca14}
\{\hat y_1,u_1\}=-2\sqrt{-\hat y_1(\hat y_1+\hat d_1)},\qquad 
\{\hat y_2,u_2\}=-2\sqrt{-\hat y_2(\hat y_2+\hat d_2)}.
\end{equation}

Defining new variables $m_i^\pm$,
\begin{equation}\label{ca15}
m_i^\pm:=1+2\,\frac{\hat y_i}{\hat d_i}\pm 2\,\sqrt{\frac{\hat y_i}{\hat d_i}
\left(\frac{\hat y_i}{\hat d_i}+1\right)},\qquad i=1,2,
\end{equation}
we finally obtain the separation equations in the form
\begin{equation}\label{ca16}
u_i^6-(H+{\cal P}b^2/2)u_i^4+\frac14((H+{\cal P}b^2)^2
-K+2a^2b^2)u_i^2-\frac{b^2\ell^2}{2}=
\frac{b^2}{4}\left({\cal P}u_i^4+a^2u_i^2-\ell^2\right)(m_i^++m_i^-).
\end{equation}
The final set of separation variables $u_i$ and $m_i^\pm$ satisfy two separation
equations above and the algebra below:
\begin{eqnarray}
\{u_1,u_2\}&=&0,\qquad \{u_j,m_k^\pm\}=0,\qquad j\neq k,\\
\{m_j^\pm,u_j\}&=&\mp 2i \,m_j^\pm,\qquad m_j^+m_j^-=1.
\end{eqnarray}
As was already mentioned, this was a result of \cite{KK87}
and \cite{KK90} in the e(3) and o(4) case, respectively.
In \cite{Ku85,KK87} these separation variables and corresponding
action variables were used for constructing the quasiclassical
spectrum of the integrals of motion of the Kowalevski top.

If we exclude the variables $m_i^+$ using the condition that $m_i^+m_i^-=1$
and if we rescale the variables $m_i^-$,
\begin{equation}\label{ca17}
m_i^+=\frac{1}{m_i^-},\qquad \widetilde m_i^-=\frac{b^2}{4}
({\cal P}u_i^4+a^2u_i^2-\ell^2)\,m_i^-,
\end{equation}
the separation equations (\ref{ca16}) will acquire the following form:
\begin{equation}\label{ca18}
\left(\widetilde m_i^-\right)^2-P_3(u_i^2)\,\widetilde m_i^-+\left( 
 P_2(u_i^2)\right)^2=0,
\end{equation}
where two polynomials $P_3(u)$ and $P_2(u)$ are
\begin{equation}\label{ca19}
P_3(u)=u^3-(H+{\cal P}b^2/2)u^2+\frac14((H+{\cal P}b^2)^2
-K+2a^2b^2)u-\frac{b^2\ell^2}{2},
\end{equation}
\begin{equation}\label{ca20}
P_2(u)=\frac{b^2}{4}\left({\cal P}u^2+a^2u-\ell^2\right).
\end{equation}
Therefore, separation equations (\ref{ca16}) amount to having 
two points, $(u_1, \widetilde m_1^-)$ 
and $(u_2, \widetilde m_2^-)$, 
on the algebraic curve $\Gamma$:
\begin{eqnarray}\label{ca21}
&&\Gamma: \qquad
m^2-P_3(u^2)\, m+\left(  P_2(u^2)\right)^2=0,\label{curve}\\
&&(u_i,\widetilde m_i^-)\in\Gamma, \qquad i=1,2.
\end{eqnarray}

The curve $\Gamma$ (\ref{curve}) is a hyperelliptic curve with the involution
$u\mapsto -u$. This 
is the curve that replaces Kowalevski's genus 2 hyperelliptic curve 
$\mu^2=R_5(\lambda)$
if one takes into account the Poisson structure and the symmetry of the problem.
The Kowalevski dynamics is therefore linearized on the corresponding hyperelliptic
Prymian of the curve $\Gamma$\footnote{ 
see the survey by Markushevich in this volume (and also his work \cite{Mar})
about interrelations between different curves for the Kowalevski top 
in the case of e(3) algebra and zero Casimir $\ell=0$}. We can bring the 
curve $\Gamma$ (\ref{ca21}) into the standard hyperelliptic form 
by shifting the variable $m$:
\begin{equation}\label{caca20}
m\mapsto \hat m=m-\frac12 \;P_3(u^2),
\end{equation}
thus getting the curve,
\begin{eqnarray}\label{caca21}
\hat m^2&=&\frac14\;\left(P_3(u^2)-2P_2(u^2)\right)
\left(P_3(u^2)+2P_2(u^2)\right)\equiv \frac14 \;R_6(u^2),\\
R_6(u)&=&u\,R_5\left(u-H/2\right),
\end{eqnarray}
with Kowalevski's polynomial $R_5(\lambda)$ given in (\ref{ke2}).

\vskip 2mm

Now we can formulate the problem: {\it to find a $2\times 2$ Lax matrix for which
the curve $\Gamma$ (\ref{ca21}) is the spectral curve or, in other words,
to find the proper $2\times2$ Lax matrix for the Kowalevski top which is related
to Kowalevski's separation of variables}.
\vskip 2mm

We must mention here the known Lax matrices for the Kowalevski top.
Let us consider only the e(3) case, as nothing is known about Lax matrices
for the o(4) case. Previous attempts to construct Lax matrices for the KT
included: (i) $2\times2$ matrix in \cite{F88}, (ii) $3\times 3$ matrix in \cite{HH}, 
(iii) $4\times 4$ and $6\times 6$ matrices in \cite{AM88}, and
(iv) $5\times 5$ (or $4\times 4$) matrix in \cite{RS}. The first three do not
respect the Poisson structure of the problem in contrast to the last one, 
which satisfies a linear $r$-matrix algebra. Unfortunately, no separation of variables
is known which can be related to the Lax matrix of \cite{RS}. In fact, as 
I said before, no separation is known which is alternative to the original 
Kowalevski's separation of the problem!

What can be guessed about the ansatz for the required $2\times2$ 
Lax matrix $T(u)$?
{}From the form of the spectral curve $\Gamma$ (\ref{ca21}) 
one can conclude that the 
entries of $T(u)$ will be polynomials in $u$ of order 
nor higher then 6, the
$\mbox{tr} \,T(u)$ will be equal to $P_3(u^2)$ and 
the $\mbox{det} \,T(u)$ will be equal
to $(P_2(u^2))^2$ and, hence, will depend only on Casimirs. 
In the next Sections we will
construct (and solve) an integrable system whose Lax matrix 
has exactly these properties.
This system will be formulated within the framework of the 
quadratic $r$-matrix algebra
for integrable systems with boundary conditions. The o(4) 
Kowalevski top, as we
will see, will correspond to a system with 
$B_2^{(1)}$-type boundary conditions.
 
%
%%%%%%%%%%%%%%%%%%%%%%%%%%%%%%%%%%%%%%%%%
%
\section{Quadratic algebra ${\cal B}$}
Starting from about 1982 the method of separation of variables 
gets connected with the $r$-matrix formalism of the quantum
inverse scattering method, developed during that
time by the Leningrad School. It was noticed by Komarov (see the 
footnote in \cite{S-old} and a full credit in \cite{Skl}) that for the $2\times 2$
$L$-operators (Lax matrices) the separation 
variables ought to be the zeros of the off-diagonal element
of the $L$-operator. This observation was fully exploited
by Sklyanin in \cite{S-old,S-Toda} who developed a
beautiful (pure algebraic) setting for the method within the framework
of the $r$-matrix technique. Since then this approach took off
and led to separations for many families of integrable systems.
Sklyanin also generalized the approach to include higher rank
$L$-operators and non-standard normalizations.
See the review \cite{Skl} where many of the examples
were exposed. An alternative, algebraic geometric approach,
which dates back to Adler and van Moerbeke \cite{AM,AM2} and 
Mumford \cite{M} and includes many researchers,
have been developed starting from about the same time
(see, for instance, \cite{Pr,HM89,AHH2,AHH,Van,Hu,Hu2}). It also led to 
many important new separations for complicated integrable 
systems and tops.
Unfortunately, we can not review this another approach,
as it would require much larger scope 
than the pure algebraic one we adopted here.

In \cite{S-old} the Goryachev-Chaplygin top and in \cite{KT89a}
the (symmetric) Neumann system and so-called Kowalevski-Chaplygin-Goryachev
top were related to special representations of the quadratic $r$-matrix algebras. 
In the present paper we show how the o(4)
Kowalevski top is related to a special representation of the quadratic
$r$-matrix algebra for integrable systems with boundary conditions.
We start by introducing the corresponding quadratic algebra ${\cal B}$.

Let $\delta\in\C$. Consider the following $L$-operator:
\begin{eqnarray}\label{ca22}
L(u)&=&\pmatrix{A(u)&B(u)\cr C(u)&D(u)},\\
A(u)&=&A_4u^4+A_3u^3+A_2u^2+A_1u+\delta,\label{AA}\\
D(u)&=&A(-u),\\
B(u)&=&B_3u^3+B_1u,\\
C(u)&=&u^5+C_3u^3+C_1u,\label{DD}\\
\mbox{deg}_uL(u)&=&\pmatrix{4&3\cr 5&4},\qquad L(-u)=L^\wedge(u).
\end{eqnarray}

Introduce $B_2$-type quadratic Poisson algebra ${\cal B}$ with eight generators,
$A_1,\ldots,A_4$, $B_1,B_3$, $C_1,C_3$, and the following Poisson brackets:
\begin{eqnarray}\label{ca23}
\{A_4,A_3\}&=&2i\,B_3,\qquad \{A_4,A_2\}=0,\qquad \{A_4,A_1\}=2i\,B_1,\\
\{A_3,A_2\}&=&-2i\,B_1,\qquad \{A_3,A_1\}=0,
\qquad \{A_2,A_1\}=2i\,(B_1C_3-B_3C_1),
\end{eqnarray}
\begin{equation}\label{ca24}
\{B_3,B_1\}=0,\qquad \{C_3,C_1\}=0,
\end{equation}
\begin{eqnarray}\label{ca25}
\{B_3,C_3\}&=&-8i\,A_3A_4,\qquad \{B_3,C_1\}=-8i\,A_1A_4,\\
\{B_1,C_3\}&=&-8i\,A_1A_4,\qquad \{B_1,C_1\}=-8i\,(-\delta A_3+A_1A_2),
\end{eqnarray}
\begin{eqnarray}\label{ca26}
\{B_3,A_4\}&=&0,\qquad \{B_3,A_3\}=4i\,B_3A_4,\\ 
\{B_3,A_2\}&=&0,\qquad\{B_3,A_1\}=4i\,B_1A_4,\\
\{B_1,A_4\}&=&0,\qquad \{B_1,A_3\}=4i\,B_1A_4,\\ \{B_1,A_2\}&=&4i\,
(B_1A_3-B_3A_1),\qquad \{B_1,A_1\}=4i\,(B_1A_2-\delta B_3),
\end{eqnarray}
\begin{eqnarray}\label{ca27}
\{C_3,A_4\}&=&4i\,A_3,\qquad \{C_3,A_3\}=4i\,(A_2-C_3A_4),\\
 \{C_3,A_2\}&=&4i\,A_1,\qquad\{C_3,A_1\}=4i\,(\delta-C_1A_4),\\
\{C_1,A_4\}&=&4i\,A_1,\qquad \{C_1,A_3\}=4i\,(\delta-C_1A_4),\\ 
\{C_1,A_2\}&=&4i\,(A_1C_3-A_3C_1),\qquad \{C_1,A_1\}=4i\,(\delta C_3-C_1 A_2).
\label{ca27-end}
\end{eqnarray}

This is a rank 2 algebra because there are four Casimirs $Q_1,\ldots,Q_4$, 
which are 
the coefficients of the $\mbox{det}\,L(u)$:
\begin{eqnarray}\label{ca28}
\det L(u)&=&Q_4u^8+Q_3u^6+Q_2u^4+Q_1u^2+\delta^2,\\
Q_4&=&A_4^2-B_3,\\
Q_3&=&2A_2A_4-A_3^2-B_3C_3-B_1,\\
Q_2&=&A_2^2+2\delta A_4-2A_1A_3-B_3C_1-B_1C_3,\\
Q_1&=&2\delta A_2-A_1^2-B_1C_1.
\end{eqnarray}

In the matrix notations, this algebra looks very compact,
\begin{equation}\label{ca29}
\{L_1(u),L_2(v)\}=[r(u-v),L_1(u)L_2(v)]+L_1(u)r(u+v)L_2(v)-L_2(v)r(u+v)L_1(u),
\end{equation}
with the $r$-matrix $r(u)$ being as follows:
\begin{equation}\label{ca30}
r(u)=\frac{-2i}{u}\pmatrix{1&0&0&0\cr 0&0&1&0\cr 0&1&0&0\cr 0&0&0&1}.
\end{equation}
In (\ref{ca29}) we use the standard notations for the tensor products,
\begin{equation}\label{ca31}
L_1(u)=L(u)\otimes \pmatrix{1&0\cr 0&1},\qquad 
L_2(v)=\pmatrix{1&0\cr 0&1}\otimes L(v).
\end{equation}
Also, in the left-hand-side of (\ref{ca29}) we have a $4\times 4$ 
matrix with the entries
$(\{L_1(u),L_2(v)\})_{ij,kl}$ $\equiv\{(L(u))_{ij},(L(v))_{kl}\}$ and 
in the right-hand-side of (\ref{ca29}) there are ($4\times4$) matrix 
commutator and products.
For more details on these notations and on quadratic $r$-matrix 
algebras see \cite{S-old,S-Toda,S88,KT89a,KT89b,Ku90,KT-Toda,KK94,KJC,Skl,Ku96,Ku97}.

The Poisson brackets for the polynomials $A(u)$, $B(u)$ and $C(u)$ 
(the polynomial
$D(u)\equiv A(-u)$) read
\begin{equation}\label{ca32}
\{A(u),A(v)\}=\frac{-2i}{u+v}\left(B(u)C(v)-B(v)C(u)\right),\quad
\{B(u),B(v)\}=0,\quad \{C(u),C(v)\}=0,
\end{equation}
\begin{equation}\label{ca33}
\{B(u),A(v)\}=\frac{-2i}{u-v}\left(B(u)A(v)-B(v)A(u)\right)+
\frac{2i}{u+v}\left(A(v)B(u)+B(v)A(-u)\right),
\end{equation}
\begin{equation}\label{ca34}
\{C(u),A(v)\}=\frac{-2i}{u-v}\left(A(u)C(v)-A(v)C(u)\right)-
\frac{2i}{u+v}\left(C(u)A(v)+A(-u)C(v)\right),
\end{equation}
\begin{equation}\label{ca35}
\{B(u),C(v)\}=\frac{-2i}{u-v}\left(A(-u)A(v)-A(-v)A(u)\right)-
\frac{2i}{u+v}\left(A(u)A(v)-A(-v)A(-u)\right).
\end{equation}

%
%%%%%%%%%%%%%%%%%%%%%%%%%%%%%%%%%%%%%%%%%
%
\section{Separation representation of the algebra ${\cal B}$}

Let us realize the algebra ${\cal B}$ (\ref{ca23})--(\ref{ca27-end}) 
in terms of 
`separation variables' $\hat u_j$, $\hat m_j^\pm$, $j=1,2$. 

First, introduce two new Poisson commuting
variables $\hat u_1$ and $\hat u_2$ as zeros of the polynomial 
$C(u)$:
\begin{equation}\label{ca41}
C(u)=u^5+C_3u^3+C_1u=u(u^2-\hat u_1^2)(u^2-\hat u_2^2),
\end{equation}
\begin{equation}\label{ca42}
\hat u_1^2+\hat u_2^2=-C_3, \qquad \hat u_1^2\hat u_2^2=C_1,
\end{equation}
\begin{equation}\label{ca43}
\{C_3,C_1\}=0\quad \Leftrightarrow \quad \{\hat u_1,\hat u_2\}=0.
\end{equation}
The `conjugated' variables $\hat m_j^\pm$ are introduced as
corresponding values of the polynomial $A(\mp u)$ at these $\hat u_j$:
\begin{equation}\label{ca44}
\hat m_j^\pm=A(\mp \hat u_j),\qquad j=1,2.
\end{equation}

It is easy to prove that these new variables obey the following 
(separated) algebra of Poisson brackets:
\begin{eqnarray}\label{ca45}
&(A)&\qquad \hat m_j^+\hat m_j^-=\det  L(\hat u_j),\\
&(B)&\qquad \{\hat m_j^\pm,\hat u_j\}=\mp 2i\,\hat m_j^\pm,\qquad j=1,2,\\
&(C)&\qquad \{\hat m_k^\pm,\hat u_j\}=\{\hat m_k^\pm,\hat m_j^\pm\}=
\{\hat u_k,\hat u_j\}=0,\qquad k\neq j.
\end{eqnarray}

Let us prove, for instance, that $\{\hat u_j,\hat m_k^-\}=-2i\,\delta_{jk}\, 
\hat m_k^-$ (cf. \cite{S-old,Ku90}). 
Notice first that by the differentiation property of the Poisson bracket,
\begin{equation}\label{ca46}
0\equiv \{C(\hat u_j),A(v)\}=C'(\hat u_j)\{\hat u_j, A(v)\}
+\{C(u),A(v)\}_{|_{u=\hat u_j}}.
\end{equation}
Hence,
\begin{equation}\label{ca47}
\{\hat u_j, A(v)\}=-\frac{1}{C'(\hat u_j)}
\,\{C(u),A(v)\}_{|_{u=\hat u_j}}.
\end{equation}
Now, substituting $u=\hat u_j$ into the algebraic relation (\ref{ca34}),
we obtain
\begin{equation}\label{ca48}
\{C(u),A(v)\}_{|_{u=\hat u_j}}=\frac{-2i}{\hat u_j-v}\,\hat m_j^-C(v)-
\frac{2i}{\hat u_j+v}\,\hat m_j^+C(v).
\end{equation}
Therefore,
\begin{equation}\label{ca49}
\{\hat u_j,\hat m_k^-\}=0,\qquad j\neq k,
\end{equation}
and
\begin{equation}\label{ca50}
\{\hat u_j,\hat m_j^-\}=-2i\,\hat m_j^-,\qquad j=1,2.
\end{equation}
The rest of the relations $(A),(B),(C)$ can be proved in the similar way.

Now we can realize the four polynomials $A(u)$, $B(u)$, $C(u)$ and $D(u)$,
whose coefficients are the generators
of the algebra ${\cal B}$,
in terms of $\hat u_j$, $\hat m_j^\pm$,
$j=1,2$, from the following data:
\begin{itemize}
\item $\quad C_3=-\hat u_1^2-\hat u_2^2$, $\quad C_1=\hat u_1^2\hat u_2^2
\quad \Rightarrow\quad C(u)$
\item $\quad A(\pm \hat u_j)=\hat m_j^\mp, \quad A(0)=\delta
\quad \Rightarrow\quad A(u)$
\item $\quad D(u)=A(-u)$
\item $\quad B(u)=\frac{A(u)D(u)-\det L(u)}{C(u)}$, $\quad C(u)$ and the numerator
have common zeros at $u=0$ and at $u=\pm \hat u_{1,2}$, so that $C(u)$ divides
the numerator.
\end{itemize}

The result is given by the formulas
\begin{eqnarray}\label{ca51}
A(u)&=&\frac{(u^2-\hat u_1^2)(u^2-\hat u_2^2)}{\hat u_1^2\hat u_2^2}\,\delta+
\frac{u(u^2-\hat u_1^2)\left((u-\hat u_2)\hat m_2^++(u+\hat u_2)\hat m_2^-\right)}
{2\hat u_2^2(\hat u_2^2-\hat u_1^2)}\nonumber\\
&&+\frac{u(u^2-\hat u_2^2)\left((u-\hat u_1)\hat m_1^++(u+\hat u_1)\hat m_1^-\right)}
{2\hat u_1^2(\hat u_1^2-\hat u_2^2)},\qquad D(u)=A(-u),
\end{eqnarray}
\begin{equation}\label{ca52}
C(u)=u(u^2-\hat u_1^2)(u^2-\hat u_2^2),\qquad B(u)=B_3u^3+B_1u,
\end{equation}
\begin{equation}\label{ca53}
B_3=A_4^2-q_4,\qquad B_1=2A_2A_4-A_3^2+(\hat u_1^2+\hat u_2^2)(A_4^2-q_4)-q_3,
\end{equation}
where $q_4$ and $q_3$ are fixed values of the Casimirs from the determinant 
(cf. (\ref{ca28})),
\begin{equation}\label{ca54}
\det L(u)=q_4u^8+q_3u^6+q_2u^4+q_1u^2+\delta^2.
\end{equation}
%
%%%%%%%%%%%%%%%%%%%%%%%%%%%%%%%%%%%%%%%%%
%
\section{$B_2^{(1)}$-type integrable system}

Let us now introduce an integrable system (of $B_2^{(1)}$-type) on 
the algebra ${\cal B}$.
Denote by $K(u)$ the following constant representation of the algebra 
(\ref{ca29})
with the $r$-matrix (\ref{ca30}):
\begin{equation}\label{ca55}
K(u)=\pmatrix{1&u\cr 0&1}.
\end{equation}
Define the integrable system by its Lax matrix $T(u)$,
\begin{equation}\label{ca56}
T(u):=K(u)L(u).
\end{equation}
Its integrals of motion are the coefficients of the trace of the Lax matrix
$T(u)$ ($\det T(u)$ is a Casimir),
\begin{equation}\label{ca57}
\mbox{tr}\, T(u)=u^6+{\cal H}_1u^4+{\cal H}_2u^2+2\delta,
\end{equation}
\begin{equation}\label{ca58}
{\cal H}_1=2A_4+C_3,\qquad {\cal H}_2=2A_2+C_1,\qquad 
\{{\cal H}_1,{\cal H}_2\}=0.
\end{equation}

The variables $\hat u_j$, $\hat m_j^\pm$, $j=1,2$, from the previous
Section {\it are} the separation variables for the integrable
system (\ref{ca58}). Indeed, because $\mbox{tr}\, T(u)=A(u)+D(u)+uC(u)$, one
obtains the following separation equations:
\begin{equation}\label{ca59}
\mbox{tr}\, T(\hat u_j)=\hat u_j^6+{\cal H}_1\hat u_j^4
+{\cal H}_2\hat u_j^2+2\delta
=\hat m_j^++\hat m_j^-,\qquad j=1,2.
\end{equation}

This is the separation of variables for the $B_2^{(1)}$-type integrable system
on the algebra ${\cal B}$ with the integrals of motion ${\cal H}_1$ 
and ${\cal H}_2$ (\ref{ca58}).

Notice here that a general constant representation of the algebra (\ref{ca29})
with the $r$-matrix (\ref{ca30}) is given by the full matrix,
\begin{equation}\label{ca60}
\hat K(u)=\pmatrix{\alpha u+\delta&\beta u\cr \gamma u&-\alpha u +\delta},
\end{equation}
which is used to generate  integrable systems with more general boundary 
conditions,
of $BC$-type (see, for instance, the examples of Toda lattice in \cite{S88} and
Kowalevski-Chaplygin-Goryachev top in \cite{KT89a}). The $D$-type boundary 
conditions for the Toda lattice are described by a non-constant matrix $\hat K(u)$ 
depending on the dynamical variables (cf. 
\cite{KT89b,Ku90,KT-Toda,KK94,KJC,Ku96,Ku97}).

%
%%%%%%%%%%%%%%%%%%%%%%%%%%%%%%%%%%%%%%%%%
%
\section{Lax matrix for the o(4) Kowalevski top}

Now we have everything to write down the $2\times2$ Lax matrix $T(u)$
for the o(4) KT. It will have the factorization (\ref{ca56}),
\begin{equation}\label{ca61}
T(u)=\pmatrix{1&u\cr 0&1}\pmatrix{A(u)&B(u)\cr C(u)& D(u)},
\end{equation}
and we only have to 
find the expressions for the polynomials $A(u)$, $B(u)$, $C(u)$ and $D(u)$
in terms of the o(4) variables $J_k$ and $x_k$, $k=1,2,3$.

Notice that two considered above integrable systems, the o(4) KT and the 
$B_2^{(1)}$-type
integrable system on the quadratic algebra ${\cal B}$, look the 
same in terms of
the separation variables, so that one can identify their 
separation variables 
and separation equations as follows:
\begin{equation}\label{ca62}
\hat u_j=u_j,\qquad \hat m_j^\pm=P_2(u_j^2)\,m_j^\pm, \qquad j=1,2,
\end{equation}
\begin{equation}\label{ca63}
P_3(u^2)=\mbox{tr}\,T(u)=u^6+{\cal H}_1u^4+{\cal H}_2u^2+2\delta,
\end{equation}
\begin{equation}\label{ca64}
\left(P_2(u^2)\right)^2=\det T(u)=q_4u^8+q_3u^6+q_2u^4+q_1u^2+\delta^2,
\end{equation}
where polynomials $P_3(u)$ and $P_2(u)$ are given in (\ref{ca19})--(\ref{ca20}).
To see that, compare the separation equations (\ref{ca59}) 
and the ones for the o(4) KT
(\ref{ca16}).

As a direct consequence, we can restore the whole algebra, i.e. get a new
representation of the algebra ${\cal B}$ in terms of the o(4) variables. 
The result is as follows:
\begin{eqnarray}\label{ca65}
A_4&=&\frac12\left( u_1^2+u_2^2-H-{\cal P}b^2/2\right)
=\frac{\hat y_1-\hat y_2}{2(u_1^2-u_2^2)}+\frac{{\cal P}b^2}{4},\\
A_2&=&\frac18\left(  (H+{\cal P}b^2)^2-K+2a^2b^2-4u_1^2u_2^2\right)
=\frac{\hat y_2u_1^2-\hat y_1u_2^2}{2(u_1^2-u_2^2)}
+\frac{a^2b^2}{4},\\
A_0&=&-\frac{b^2\ell^2}{4},\qquad A_3=\frac{i}{8}\{ H,u_1^2+u_2^2 \},\qquad
A_1=-\frac{i}{8}\{ H,u_1^2u_2^2 \},\\
C_3&=&-u_1^2-u_2^2,\qquad C_1=u_1^2u_2^2,\\
B_3&=&\frac14 (u_1^2+u_2^2-H)(u_1^2+u_2^2-H-{\cal P}b^2)
=\frac{\hat y_1-\hat y_2}{4(u_1^2-u_2^2)}
\left( \frac{\hat y_1-\hat y_2}{u_1^2-u_2^2}+{\cal P} b^2\right),\\
B_1&=&2A_2A_4-A_3^2+(u_1^2+u_2^2)B_3-{\cal P} a^2b^4/8.
\end{eqnarray}
Or, explicitly,
\begin{eqnarray}\label{ca66}
A_4&=&-\frac12\left( X^2+J_3^2-{\cal P}b^2/2\right),\\
A_2&=&\frac{J^2}{2}\left( X^2+J_3^2 \right)+bx_2J_3X-bx_1J_3^2+\frac{b^2}{2}
\left(  {\cal P}J_3^2-x_3^2+\frac{a^2}{2}\right),\\
A_0&=&-\frac{b^2\ell^2}{4},\qquad
A_3=-\frac{i}{2}\left( X^3 +(J_3^2+bx_1-{\cal P}b^2)X+bx_2J_3 \right),\\
A_1&=&\frac{i}{2}\left(  J^2X^3+2bx_2J_3X^2
+\left( J_3^2(J^2-bx_1)+b^2(x_2^2-x_3^2
-{\cal P}J_1^2) \right.\right.\\
&&\left.\left. +bJ_1(x_1J_1+x_2J_2)\right)X+bx_2J_3(J_2^2+J_3^2-bx_1)+J_1J_2J_3
(bx_1-{\cal P}b^2)\right),\nonumber\\
C_3&=&X^2-J^2+2bx_1-{\cal P}b^2,\\
C_1&=&-J^2X^2-2b(x_2J_3-x_3J_2)X-2b\ell J_1-b^2(x_2^2+x_3^2-{\cal P}J_1^2),\\
B_3&=&\frac14 \left(X^2+J_3^2\right)\left(X^2+J_3^2-{\cal P}b^2\right),\\
B_1&=&-\frac14\left(  J^2X^4+2bx_2J_3X^3
+\left(2J_3^2(J^2-bx_1+{\cal P}b^2/2)+a^2b^2-b^2(2x_3^2+x_1^2)
\right)X^2\qquad\right.\\
&&\left. +2bx_2J_3(J_3^2-bx_1)X+J_3^4(J^2-2bx_1)
-b^2J_3^2(x_3^2-x_1^2+{\cal P}(J_1^2+J_2^2))
+{\cal P}b^4x_3^2\right),\quad\nonumber
\end{eqnarray}
where
\begin{equation}\label{ca67}
X:=\frac{J_1J_3+bx_3}{J_2},\qquad J^2:=J_1^2+J_2^2+J_3^2.
\end{equation}
Recall that 
\begin{equation}\label{ca68}
\ell=x_1J_1+x_2J_2+x_3J_3,\qquad a^2=x_1^2+x_2^2+x_3^2-{\cal P}J^2.
\end{equation}

The formulas (\ref{ca66})--(\ref{ca68}) give the polynomials 
$A(u)$, $B(u)$, $C(u)$
and $D(u)$ (cf. (\ref{AA})--(\ref{DD})), which together with the formula
(\ref{ca61}) define the $2\times 2$ Lax matrix $T(u)$ for the o(4)
Kowalevski top. This Lax matrix satisfies the quadratic ($r$-matrix) algebra,
corresponding to the $B_2$-type boundary conditions, and it is the one
which is responsible for the separation of variables for the Kowalevski top.

Recall that the spectral curve $\Gamma$ of the Lax matrix $T(u)$ (\ref{ca61})
has the following form: 
\begin{equation}\label{ca69}
\Gamma:\quad \det(T(u)-m)\equiv m^2-\mbox{tr}\,T(u)\;m+\det T(u)=0,
\end{equation}
\begin{equation}\label{ca70}
\mbox{tr}\,T(u)=P_3(u^2)=u^6-(H+{\cal P}b^2/2)u^4+\frac14((H+{\cal P}b^2)^2
-K+2a^2b^2)u^2-\frac{b^2\ell^2}{2},
\end{equation}
\begin{equation}\label{ca71}
\det T(u)=\left(P_2(u^2)\right)^2=\frac{b^4}{16}\left({\cal P}u^4+a^2u^2-\ell^2\right)^2.
\end{equation}

Knowing the $r$-matrix algebraic structure (\ref{ca29})--(\ref{ca30})
of the found Lax matrix, it is not difficult to derive the full Lax pair, namely: 
the Hamiltonian flow given by the Kowalevski Hamiltonian $H$ (cf. (\ref{dot})),
\begin{equation}\label{ca72}
\dot{J_1}=J_2J_3,\qquad \dot{J_2}=-J_3J_1-bx_3,\qquad \dot{J_3}=bx_2,
\end{equation}
\begin{equation}\label{ca73}
\dot{x}_1=2x_2J_3-x_3J_2,\qquad \dot{x}_2=x_3J_1-2x_1J_3+b{\cal P}J_3,
\qquad \dot{x}_3=x_1J_2-x_2J_1-b{\cal P}J_2,
\end{equation}
has the following Lax pair, $T(u)$ and $M(u)$,
\begin{equation}\label{ca74}
\dot{T}(u)=-i[T(u),M(u)],\qquad M(u)=\pmatrix{u & 2A_4 \cr 2 & -u}.
\end{equation}

%
%%%%%%%%%%%%%%%%%%%%%%%%%%%%%%%%%%%%%%%%%
%
\section*{Acknowledgements}

I want to thank Igor Komarov, Evgueni Sklyanin and Pol Vanhaecke for interesting
discussions of the Kowalevski top. I acknowledge the support of the 
EPSRC.
%\pagebreak

%
%%%%%%%%%%%%%%%%%%%%%%%%%%%%%%%%%%%%%%%%%
%
\def\cprime{$'$}

\end{document}